# Spin-order driven Fermi surface revealed by quantum oscillations in an underdoped high $T_\mathrm{c}$ superconductor

Suchitra E. Sebastian[1], N. Harrison[2], C. H. Mielke[2], Ruixing Liang[3,4], D. A. Bonn[3,4], W. N. Hardy[3,4] & G. G. Lonzarich[1]

We use quantum oscillation measurements to distinguish between spin and orbital components of the lowest energy quasiparticle excitations in $YBa_2Cu_3O_{6.54}$, each of which couple differently to a magnetic field. Our measurements reveal the phase of the observed quantum oscillations to remain uninverted over a wide angular range, indicating that the twofold spin degeneracy of the Landau levels is virtually unaltered by the magnetic field. The inferred suppression of the spin degrees of freedom indicates a spin density-wave is responsible for creation of the small Fermi surface pockets in underdoped $YBa_2Cu_3O_{6+x}$—further suggesting that excitations of this phase are important contributors to the unconventional superconducting pairing mechanism.

[1] *Cavendish Laboratory, Cambridge University, JJ Thomson Avenue, Cambridge CB3 0HE, U.K.*

[2] *NHMFL, Los Alamos National Laboratory, MS E536, Los Alamos, NM 87545, U.S.A.*

[3] *Department of Physics and Astronomy, University of British Columbia, Vancouver V6T 1Z4, Canada*

[4] *Canadian Institute for Advanced Research, Toronto M5G 1Z8, Canada*

Potential mechanisms of high-$T_c$ superconductivity may be elucidated by identifying precursory competing phases, the low energy excitations of which may bind Cooper pairs. The different characteristics of these contending phases must ultimately be reflected in differences in orbital, spin or charge quantization at the Fermi surface, providing a simple means for them to be distinguished[1-5]. We use magnetic field orientation dependent quantum oscillation measurements[6] to distinguish between spin and orbital components of the quantized Landau levels in $YBa_2Cu_3O_{6.54}$.

We present angle-dependent quantum oscillations measured in the underdoped cuprate $YBa_2Cu_3O_{6.54}$ (nominal hole doping $p \approx 0.10$ [Ref. 9] for which the observed small Fermi surface pocket occupies about 2% of the unreconstructed Brillouin zone area (Fig. 1), similar to the α pocket observed in other underdoped members of the $YBa_2Cu_3O_{6+x}$ family[10-15]. The angular dependence of the quantum oscillations associated with this pocket (Fig. 2) confirm it to be a corrugated cylinder typical of quasi-two-dimensional electronic structures in layered metals[16,17]. The anisotropy of the electronic structure of this family of materials accentuates the sensitivity of Landau quantized states to the orientation of **H**, facilitating the use of quantum oscillations to distinguish between orbital, charge, and spin quantum numbers[6,7,8].

We look for interference effects in the angular dependence of the quantum oscillation amplitude measured in $YBa_2Cu_3O_{6.54}$. In the case where the spin moment is unaffected by Fermi surface reconstruction, interference between the Zeeman split spin components invariably leads to a 'spin damping' factor $R_\mathrm{s} = \cos(\phi/2)|_\mathrm{orbital}$ that modifies the orbitally averaged amplitude ($A_\theta$) of the fundamental component of the quantum oscillations[P] ($\phi$ is the phase difference between spin up and spin down quantum oscillations). The standard signature of the consequent angular oscillations in $A_\theta$ is a series of special "spin zero" angles at which $R_s = 0$ caused by destructive interference between the two opposite spin components[6]. Importantly, the observation of such an interference effect (i.e. $R_s \neq 1$) in the angular dependence of the quantum oscillation amplitude occurs only where spin up and down are individually well defined, resulting in an effective value of $\gamma \sim 1$ for the spin moment (associated with the relative population of spin-up and –down Landau levels in a magnetic field) close to the free electron.

A striking finding from our current measurements of underdoped $YBa_2Cu_3O_{6.54}$ is that there are **no** spin zeros over the wide angular range accessed from -52° to 57°: on the contrary, the oscillations retain approximately the same phase over the entire range (Fig. 3B). The absence of spin zeros over this wide angular range in $YBa_2Cu_3O_{6.54}$ is unsupported by a value of $\gamma \sim 1$ for the measured Fermi surface sheet (see Fig. 3). This is in contrast to the value of $\gamma$ in the cuprates (considered to be an average over the entire Fermi surface comprising multiple sheets), which is given by the measured Wilson ratio that appears in the expression for the enhancement in spin susceptibility

$$\frac{\chi^\mathrm{spin}}{\chi^\mathrm{spin}_\mathrm{electron}} = \gamma \frac{m*}{m_\mathrm{e}}\bigg|_\mathrm{FS}.$$ Given the proximity to an antiferromagnetic instability, the Wilson

ratio is expected to be close to 1 in the cuprates, and thermodynamic as well as nuclear magnetic resonance spectroscopy studies report a value of $\gamma \sim 1.1$ in YBa$_2$Cu$_3$O$_{7-x}$ [Refs. 18-21].

Our measurements therefore indicate a suppression of the effective spin moment on the measured Fermi surface sheet ($\alpha$) as a consequence of Fermi surface reconstruction. Moreover, we determine the expected location of spin zeroes for YBa$_2$Cu$_3$O$_{6.54}$ to confirm that were they present, they would have been accessed within the angular range -52° to 57°. To first order, the ratio of the phase difference between spins ($\phi$) to that between consecutive quantum oscillations ($2\pi$) is equivalent to twice the ratio of the Zeeman splitting[18,19] ($\Delta\varepsilon = \pm\gamma\mu_B B$)— reflecting the ratio of the spin quantisation to the cyclotron energy ($\hbar\omega_c = \hbar e B_\perp / m_0^*$; $B_\perp = B\cos\theta$) in the orbital quantisation. The location of the spin zeroes is therefore given by $\frac{2\gamma m_0^*}{m_e \cos\theta} = 2n+1$, where $n$ is an integer[6]. In the case of underdoped YBa$_2$Cu$_3$O$_{6+x}$, were $\gamma \sim 1.1$ for the measured section of Fermi surface, we would expect the quantum oscillation amplitude to exhibit a spin zero at $|\theta| \approx 42°$ in Fig. 3A, accompanied by an inversion in the phase of quantum oscillations beyond this angle for the experimentally estimated cyclotron mass of $m_0^* = 1.7 \pm 0.1$ $m_e$ [Fig. 2A]. The effect of spin-orbit coupling is neglected here for simplicity, since it has a relatively small effect in YBa$_2$Cu$_3$O$_{6+x}$, with a measured Landé g-factor $\sim 2.1$ [ref. 18] – a value that is also close to the free-electron value of 2 [ref. 19].

The approximately constant phase of the quantum oscillations we observe between -52° and 57° in YBa$_2$Cu$_3$O$_{6.54}$ therefore reveals a vanishingly small value of $\gamma_\alpha$ in Eqn. (1), corresponding to the measured small $\alpha$ Fermi surface pockets (inferred from Fig. 3; also indicated in the Figure is a small range of very low $\gamma$ that cannot be ruled out). The absence of spin zeroes is therefore inconsistent with small Fermi surface pockets arising from within the band-structure (where $\gamma$ close to unity[1-5,22-25]) or pockets resulting from an order parameter that couples states of like spin (Fig. 4) – for example, forms of charge ordering. An example is shown in the inset to Fig. 3B – for the charge-density wave material α-(BEDT-TTF)$_2$KHg(SCN)$_4$, which has very similar Fermi surface parameters (i.e. frequency, effective mass and Wilson ratio) to YBa$_2$Cu$_3$O$_{6.54}$. As expected, a 'spin zero' accompanied by phase inversion of the oscillations is observed experimentally at an angle of ~ 43° [Ref. 26].

Our finding that the effective moment of the spin doublet is suppressed for the measured orbit (i.e. $\gamma_\alpha \rightarrow 0$), has implications for the nature of ordering associated with the creation of the observed $\alpha$ Fermi surface sheet in underdoped YBa$_2$Cu$_3$O$_{6+x}$. For types of magnetic ordering where opposite spins couple, a suppression of the effective moment of the spin doublet can result (i.e. $\gamma \rightarrow 0$), in which case $B$ cannot lift the Landau level spin degeneracy to leading order. The order parameter consistent with *both* the creation of small Fermi surface pockets and the coupling of spin up and down states is a spin-density wave state[27]. In this case, spin is no longer well defined for the Landau quantized states, so that the effective moment of the spin doublet becomes $\gamma \sim 0$ in Eqn. (1) (see Fig. 3c). The layered Bechgaard salts[27,29] provide well known examples of spin-density-wave phases exhibiting such a robust twofold degeneracy of the Landau levels, where $\gamma \sim 0$ is rigorously confirmed by experiment. In Fig. 3C, we also compare the $\theta$-dependent attenuation of the quantum oscillation amplitude in $YBa_2Cu_3O_{6.54}$ with that predicted for $\gamma = 0$ (assuming the exponential damping factor due to disorder scattering depends only on $B_\perp$), and find reasonable consistency.

Our present finding that $\gamma_\alpha \ll 1$ indicates that spin ordering is associated with the creation the $\alpha$ Fermi surface pockets in $YBa_2Cu_3O_{6.54}$. Additional evidence is provided by recent elastic neutron scattering measurements made on nearby $YBa_2Cu_3O_{6+x}$ sample compositions that reveal magnetic field-enhancement of spin ordering[30-32]. The current findings are therefore likely applicable to a range of hole dopings within the underdoped regime[15]. While other order parameters such as a triplet $d$-density wave (also referred to as "unconventional spin-density wave")[28] can also result in a suppression of $\gamma$, there has to date been no evidence for such forms of order in the cuprates.

Our results indicate that at measured magnetic fields above the order of 30 T, the ground state associated with the measured small Fermi surface pockets in the cuprates is a spin-density wave. Since the present technique provides direct access to the lowest energy excitations[1-5], the proximity to spin ordering and its sensitivity to a magnetic field indicates the existence of soft spin excitations in $YBa_2Cu_3O_{6.54}$ and potentially other dopings in the underdoped regime of high temperature superconductors[15,30-33]. An implication is that magnetic fluctuations may play an important role in the pairing mechanism of carriers in cuprate superconductors. These observations contribute to an overarching picture of the intimate connection of spin excitations to unconventional superconducting pairing.

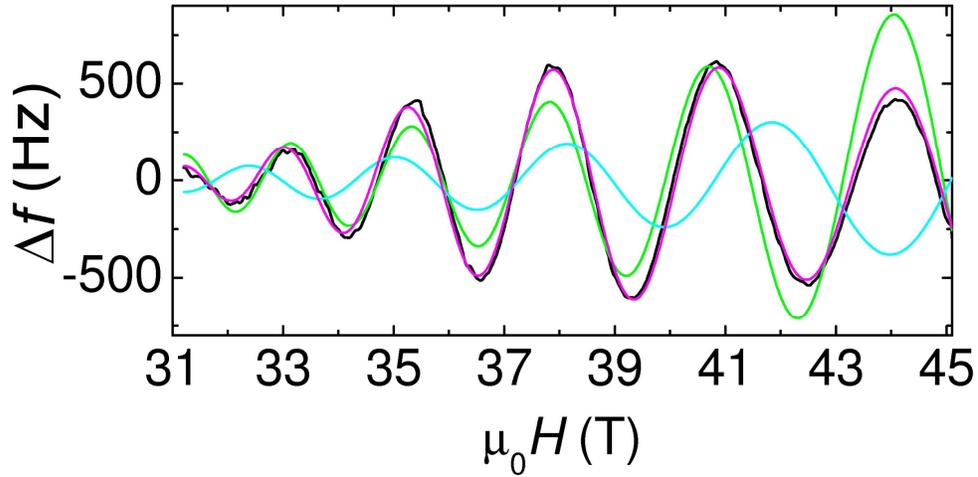

**Fig. 1. Magnetic quantum oscillations in YBa$_2$Cu$_3$O$_{6.54}$.** The black line shows quantum oscillations in the resonance frequency (*f*) of a tunnel diode oscillator (TDO) to which the sample is coupled inductively, which is proportional to the change in sample in-plane resistivity (see Appendix). For clarity, only the frequency shift ($\Delta f$) is shown. The beat structure suggests two closely spaced frequencies. A two-frequency fit (shown by the purple line) to

$$R_T \left( a_{\text{belly}} \cos[2\pi F_{\text{belly}}/B + \phi_{\text{belly}}] + a_{\text{neck}} \cos[2\pi F_{\text{neck}}/B + \phi_{\text{neck}}] \right) B^{1/2} \exp(-\Gamma/B_\perp)$$

(limited by the short range in *B* and large damping factors), yields a prominent higher frequency $F_{\text{belly}} = 530 \pm 10$ T (shown by the green line) and a lower frequency $F_{\text{neck}} = 430 \pm 10$ T (of approximately one third the amplitude, shown by the cyan line) similar to that reported in ref. 16, with damping factor $\Gamma = 240 \pm 40$ T). $F_{\text{belly}}$ is consistent with the "belly frequency" of a warped corrugated cylindrical Fermi surface produced by the dispersion

$$\varepsilon_k \approx \frac{\hbar^2 k_\parallel^2}{2m_0^*} - 2t_c \cos(k_z c), \text{ where } k_\parallel = \sqrt{k_x^2 + k_y^2}$$ and *c* is the interlayer lattice spacing.

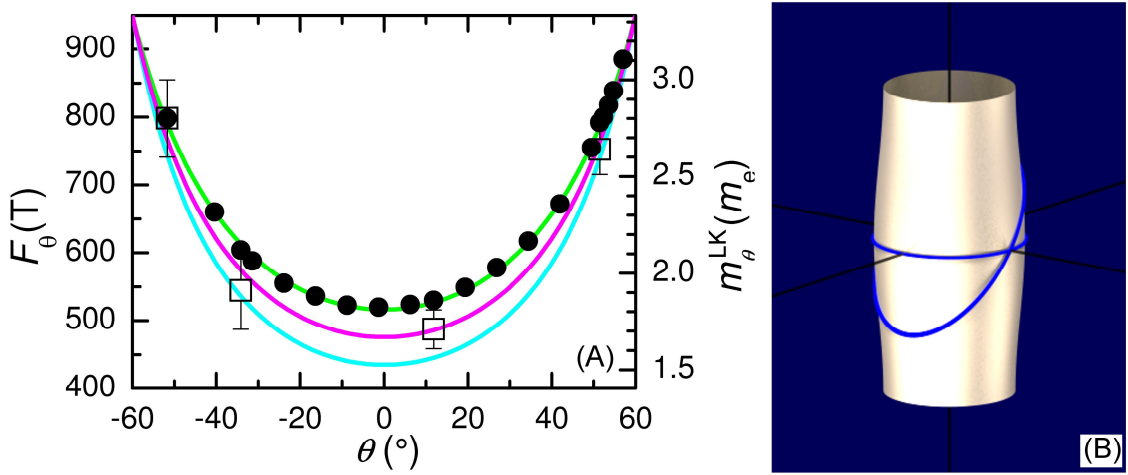

**Fig. 2**, **Fermi surface topology.** A, Magnetic field orientation-dependence of the dominant higher quantum oscillation Fourier frequency $F_{neck,\theta}$ (circles and left axis) as determined from a fast Fourier transform (FFT) of the measured oscillations (in which the lower, weaker-amplitude frequency[16] is unresolved (see Appendix)). $F_{belly,\theta}$ is related to the orientation-dependent Fermi surface cross-sectional area $A_k$ in $k$-space by the Onsager relation $F_\theta = (\hbar/2\pi e)A_k$ (averaged over spin). A fit (green line) to the expected angular form

$$F_{belly,\theta}\cos\theta = F_0 + \Delta F\, J_0\!\left(\pi\bar{k}_F \tan\theta/k_c\right),$$

where $k_c = \pi/c$ and $\bar{k}_F = \sqrt{2eF_0/\hbar}$ is the calliper radius of the Fermi cylinder[17], yields $F_0 = 480 \pm 10$ T, $\Delta F = 40 \pm 10$ T and $|t_c| \approx 1.0 \pm 0.3$ meV. The slightly lower estimate for $F_{belly,0} \approx 520 \pm 10$ T compared to Fig. 1 is likely due to a small admixing of $F_{neck}$ in the FFT. The cyan line shows the implicit minimum cross-section 'neck frequency' $F_{neck} = F_0 - \Delta F$ (consistent with the lower frequency in (Fig. 1)). Also plotted (squares and right axis), is the experimentally determined effective mass obtained by performing fits of the temperature-dependent quantum oscillation amplitude to $A_T = A_0 R_T = A_0 X_{T,\theta} / \sinh X_{T,\theta}$ at different angles, where $X_{T,\theta} = 2\pi^2 k_B m_\theta^* T / \hbar e B$ (Appendix). The $\theta$-dependent effective mass is further fit by $m_\theta^* \approx m_0^* / \cos\theta$ (magenta line) yielding $m_0^*$ = 1.7 ± 0.1 $m_e$. **B**, Corrugated cylindrical Fermi surface model consistent with the measurements in (A), with schematic orbits shown at $\theta$ = 0° and 50°.

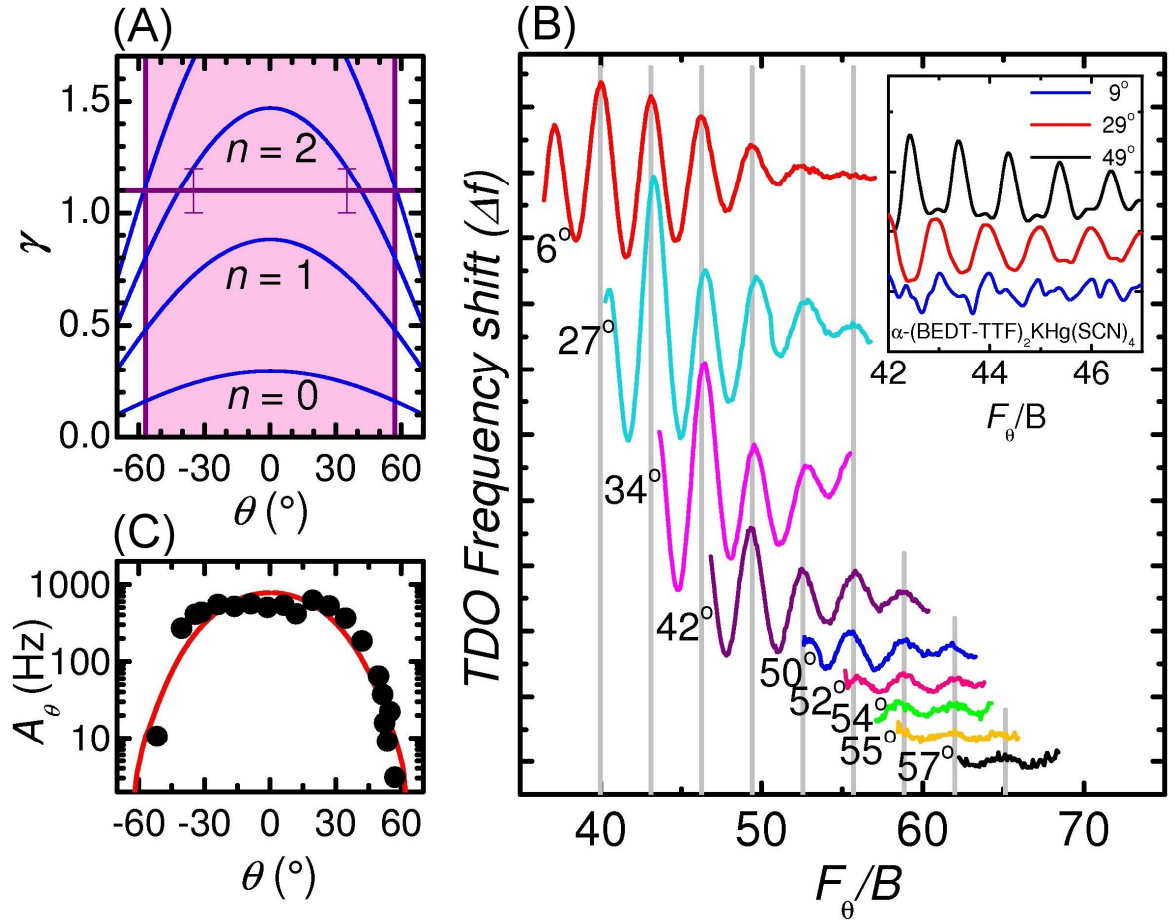
Fig. 3. **Magnetic field orientation-dependence of the quantum oscillations in YBa$_2$Cu$_3$O$_{6.54}$.** **A**, A plot of anticipated "spin zero" angles (blue curves) as a function of $\gamma$ determined by setting $\cos\left(\pi\gamma\frac{m_0^*}{m_e}\frac{B}{B_\perp}\right) = 0$, so that the angle of the $n^{th}$ zero is given by $2\gamma m_0^* = (2n+1)\,m_e\cos\theta$. For a Wilson ratio $\gamma = 1.1$ [ref. 20], given the experimental $m_0^* = 1.7\,m_e$, the first spin zero would be expected to occur at $|\theta| \approx 42°$. B, Quantum oscillations plotted versus the Landau level index $\nu = F_\theta/B$ (offset for clarity after polynomial background subtraction) shows that the relative phase of the oscillations is invariant for successive values of $\theta$. The absence of spin-zeros suggests that $\gamma$ is vanishingly small (the only finite values of $\gamma$ that cannot be ruled out are in the low range $0.3 < \gamma < 0.5$). The inset shows the contrasting case of a charge density wave material $\alpha$-(BEDT-TTF)$_2$KHg(SCN)$_4$, where a 'spin zero' accompanied by phase inversion of the oscillations is observed experimentally at an angle $\approx 43^\circ$ [Ref. 26]. **C**, Angular dependent amplitude of $F_{\text{belly}}$ at $B \approx 40$ T, extracted from the data using fits at each angle to

$$R_T \left(a_{\text{belly}} \cos[2\pi F_{\text{belly}}/B + \phi_{\text{belly}}] + a_{\text{neck}} \cos[2\pi F_{\text{neck}}/B + \phi_{\text{neck}}]\right) B^{1/2} \exp(-\Gamma/B_\perp)$$

where $F_{\text{belly}} \cos\theta = F_0 + \Delta F \, J_0\left(\pi \bar{k}_F \tan\theta / k_c\right)$ and $F_{\text{neck}} \cos\theta = F_0 - \Delta F \, J_0\left(\pi \bar{k}_F \tan\theta / k_c\right)$

with invariant $F_0$ and $\Delta F$. The red line shows the anticipated orientation-dependent amplitude according to $a'_{\text{belly}} R_T R_S B^{1/2} \exp(-\Gamma/B_\perp)$ at 40 T (i.e. where $\gamma = 0$, and only $a'_{\text{belly}}$ and $\Gamma$ are adjustable parameters), yielding reasonable agreement with experimental amplitude (Appendix).

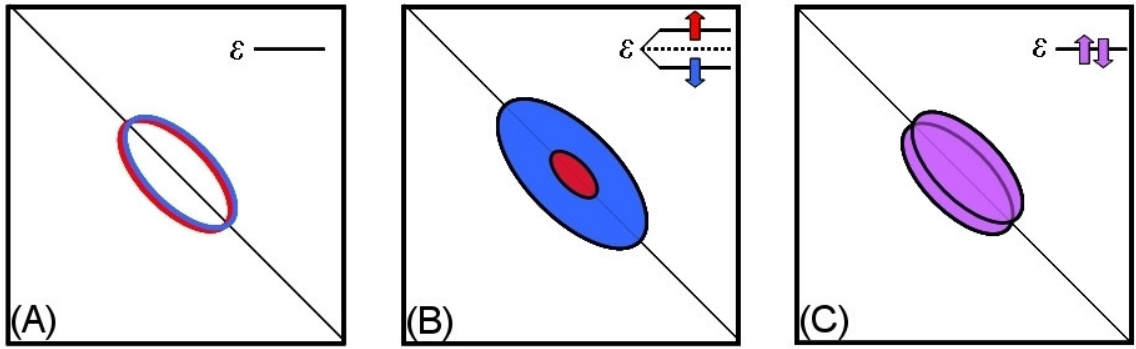

**Fig. 4. Example of the effect of Zeeman splitting on Fermi surface reconstruction. A**, Schematic of a two-fold degenerate hole-like Fermi surface pocket in the absence of a magnetic field, intersected by a Bragg reflection plane (diagonal line). **B**, Schematic showing the Zeeman splitting of the twofold degeneracy (if spin up and spin down are individually well defined so that $\gamma = 1$). The red area denotes a spin up Fermi surface while the blue area denotes a spin down Fermi surface. **C**, Schematic showing how the pocket size remains largely invariant to Zeeman splitting if the order parameter couples up and down spins (as in the case of a spin-density wave, so that $\gamma = 0$). The purple areas show Fermi surfaces comprising spins where the effective moment has been suppressed, potentially due to spin flips at Bragg reflection.

## *Appendix*

## Materials

Detwinned single crystals of YBa$_2$Cu$_3$O$_{6.54}$ of dimensions $0.5 \times 0.8 \times 0.1$ mm$^3$ were grown and prepared at the University of British Columbia[A1]. The samples are attached to a coil of ~ 5 turns that forms part of a tunnel diode oscillator (TDO) circuit[A2], resonating at ~ 46 MHz when the sample is superconducting and dropping by ~ 50 KHz on crossing over into the high magnetic field resistive state. The coil and sample are rotated in situ. The quantum oscillations occur by way of the Shubnikov-de Haas effect, which cause the skin depth and resonance frequency to oscillate in reciprocal magnetic field. Magnetic fields of up to 45.1 T are provided by the hybrid magnet at the National High Magnetic Field Laboratory in Tallahassee, with temperatures between ~ 0.5 K and 4.2 K provided by an in-house $^3$He refrigerator.

## Frequency and effective mass analysis

Fourier analysis of the quantum oscillations resolves the higher prominent frequency of $F_{\text{belly},0} = 520 \pm 10$ T over the limited data range (see Fig. A1). The cyclotron mass $m_\theta^*$ is determined at several different angles $\theta$ by fitting the temperature dependent quantum oscillation amplitude to the standard Lifshitz-Kosevich factor[A3] $A_T = A_0 \, X_{T,\theta} / \sinh X_{T,\theta}$, where $X_{T,\theta} = 2\pi^2 k_B m_\theta^* T / \hbar e B$, shown in Fig. A2.

**Angular dependent quantum oscillation amplitude**

Shubnikov-de Haas oscillations originating from a quasi-two-dimensional Fermi surface of the form depicted in Fig. 1c are given by

$$a_\theta = a_0 \cos\left(\frac{2\pi F_0}{B_\perp} + \varphi\right) \left.\cos\left(\pi\gamma \frac{m_0^*}{m_e}\frac{B}{B_\perp}\right)\right|_{\text{orbital}} \exp\left(-\frac{\Gamma}{B_\perp}\right) \times \\ J_0\left(\frac{2\pi\Delta F\, J_0\left(\pi\bar{k}_F \tan\theta/k_c\right)}{B}\right) R_T \quad , \quad (A1)$$

where $J_0(x)$ is a zeroth order Bessel function[A3-A5]. This can be expanded (in the limit of small $\frac{B}{2\pi\Delta F}$) as

$$a_\theta = \left(a_{\text{belly}} \cos\left(\frac{2\pi F_{\text{belly}}}{B} + \varphi_{\text{belly}}\right) + a_{\text{neck}} \cos\left(\frac{2\pi F_{\text{neck}}}{B} + \varphi_{\text{neck}}\right)\right) \times \\ \left.\cos\left(\pi\gamma \frac{m_0^*}{m_e}\frac{B}{B_\perp}\right)\right|_{\text{orbital}} \exp\left(-\frac{\Gamma}{B_\perp}\right) f(T,B,\theta) \quad , \quad (A2)$$

where $f(T,B,\theta) \approx R_T \left(B/\Delta F\, J_0\left(\pi\bar{k}_F \tan\theta/k_c\right)\right)^{1/2}/\pi$, providing justification for the two-frequency expression used to interpolate the quantum oscillation amplitude at ≈ 40 T in Fig. 3.

Since $f(T,B,\theta) \propto B^{1/2}$, the magnetic field dependence of the amplitude

$$A_\theta = A_0 \cos\left(\pi\gamma \frac{m_0^*}{m_e}\frac{B}{B_\perp}\right)\bigg|_{\text{orbital}} \exp\left(-\frac{\Gamma}{B_\perp}\right) f(T,B,\theta) \quad (A3)$$

of each of the "belly" and "neck" frequencies is similar to that of a maximal or minimal Fermi surface cross-section according to the Lifshitz-Kosevich theory[A3]. The introduction of a second nearest layer hopping term, $t_{2c}$, would change the respective curvature factors at the extremal "belly" and "neck" parts of the corrugated cylindrical Fermi surface[A3], leading to differences in $a_{belly}$ and $a_{neck}$. A two-frequency fit to the experimental data presented in Fig. 1c yields $a_{belly} \sim 3 a_{neck}$, potentially implying that $t_{2c} \sim -0.2 t_c$.

**Wilson ratio**

The Wilson ratio is given by $\gamma$ in the equation $\dfrac{\chi^{spin}}{\chi^{spin}_{electron}} = \gamma \dfrac{m^*}{m_e}\bigg|_{FS}$. Typically, $\gamma \geq 1$ in most conventional metals[A3], whereas $\gamma \gg 1$ close to a ferromagnetic instability[A7]. Close to an insulating antiferromagnetic instability, $\gamma$ is predicted to be of order 1 as the strength of the correlations is increased[A8]. It is worth noting that $\gamma \sim 1$ is not unique to the cuprates[A9,A10], but is found to be true in wide variety of layered systems including some layered ruthenates and organic conductors[A11].

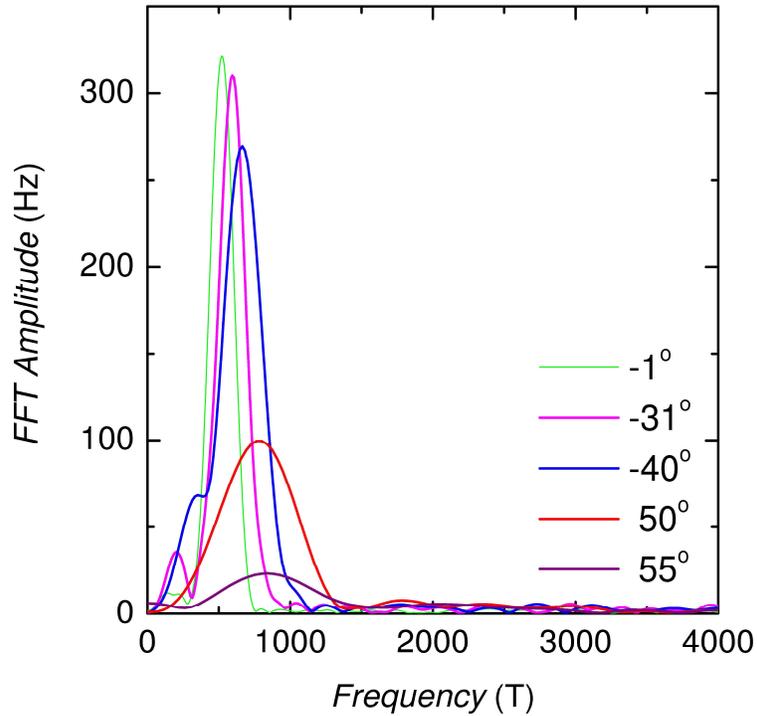

**Fig. A1. Example Fourier transforms of the quantum oscillations in YBa$_2$Cu$_3$O$_{6.54}$.** Fourier analysis, in which the oscillations are multiplied by a Hanning window function after background subtraction, yield a single prominent peak at all orientations $\theta$. Fits to the beat structure in Fig. 3B reveal the Fourier amplitude to be dominated by the belly frequency $F_{belly}$ of a corrugated cylindrical Fermi surface, yielding the angular dependence of the Fourier frequency plotted in Fig. 2A.

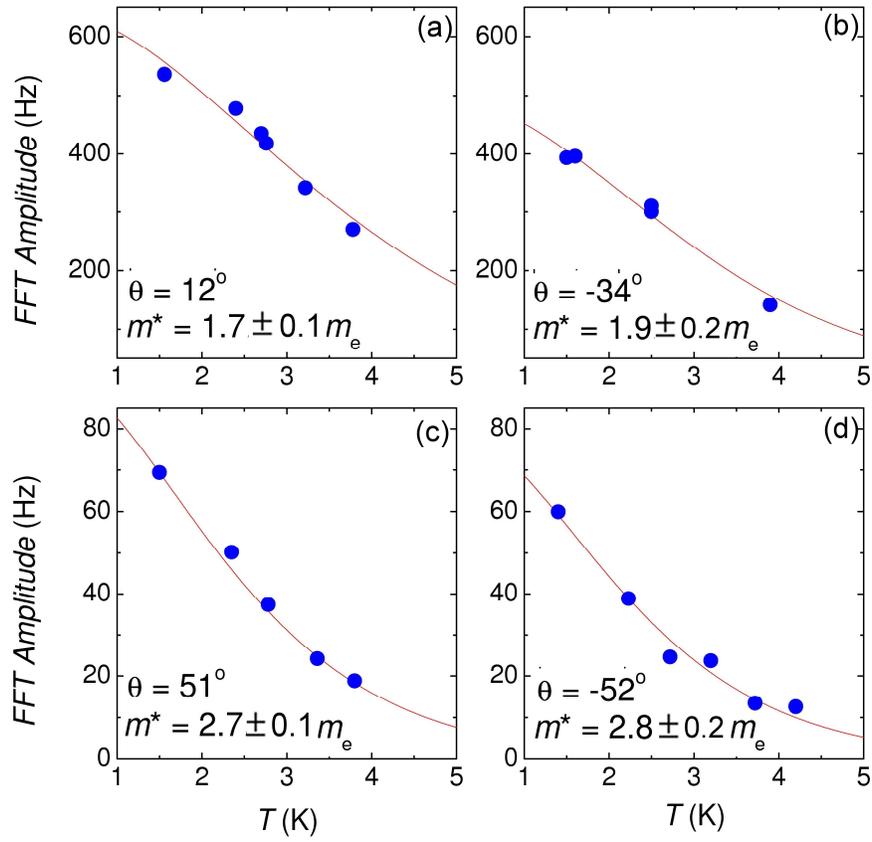

**Fig. A2. Fits to temperature dependent quantum oscillation data.** The data and LIfshitz-Kosevitch fits are shown together with the fitted effective masses $m^*_\theta$ and the angles $\theta$ at which each measurement was made.